\documentclass[twocolumn,showpacs,preprintnumbers,amsmath,amssymb]{revtex4-1}

\usepackage{graphicx}
\usepackage{amssymb}
\usepackage{hyperref}

\begin{document}

\title{From magnetic to Fermi Liquid behavior in CeCo$_{1-x}$Fe$_x$Si alloys}

\author{J.G. Sereni, M. Gomez Berisso, D. Betancourth, V.F. Correa}

\address{Low Temperature Division, CAB-CNEA, CONICET, 8400 S.C. de Bariloche, Argentina}

\author{N. Caroca Canales, C. Geibel}

\address{Max-Planck Institute for Chemical Physics of Solids, D-01187
Dresden, Germany}

\date{\today}

\begin{abstract}

{Structural, magnetic and thermal measurements performed on
CeCo$_{1-x}$Fe$_x$Si alloys are reported. Three regions can be
recognized: i) Co-rich ($x\leq 0.20$) with a decreasing long range
antiferromagnetic order which vanishes at finite temperature, ii)
an intermediate region ($0.20< x\leq 0.30$) showing a broad
magnetic anomaly ($C_A$) in specific heat and iii) the
non-magnetic region progressively changing from a non-Fermi-liquid
type behavior towards a Fermi liquid one as Fe concentration
increases. The $C_A$ anomaly emerges as an incipient contribution
above $T_N$ already at $x=0.10$, which indicates that this
contribution is related to short range correlations likely of
quasi-two dimensional type. Both, $T_N$ transition and $C_A$
anomaly are practically not affected by applied magnetic field up
to $B\approx 10$ Tesla.}

\end{abstract}


\maketitle

\section{Introduction}

Magnetic correlations play a basic role in systems accessing to
phase transformations \cite{Jongh}. Besides canonical thermal
fluctuations driving a 3-dimensional (3D) magnetic system into a
long range magnetic order (LR-MO) ground state (GS), there are
different types of fluctuations allowing to explore alternative
minima for its free energy. For example, novel exotic phases may
occur due to geometrical frustration as alternative GS
\cite{Toulouse}. Low dimensionality is another factor that
enhances fluctuations due to geometrical constraints on the
propagation of the order parameter.

Magnetic interactions in intermetallic compounds are mainly driven
by the well know RKKY mechanism, which is essentially of 3D
character. Although real low dimensionality is unlikely in
intermetallic compounds, strongly anisotropic structures favor
similar effects like in those systems resembling multi-layer
structures \cite{Venturini}. Among Ce equi-atomic ternaries,
CeFeSi and CeScSi type structures provide the possibility to
explore that alternative. The tetragonal CeFeSi-type structure
builds up from two consecutive square planes (with W-type
configuration) stacked up along 'c' direction with BaAl$_4$
blocks, following the Ce-Ce-Si-Fe$_2$-Si-Ce-Ce sequence.

In this work we report on structural, magnetic and thermal
properties of CeCo$_{1-x}$Fe$_x$Si alloys investigated all along
the concentration range. The respective stoichiometric limits are
antiferromagnetic with localized $4f$ moments (CeCoSi, $T_N =
8.8$\,K \cite{Cheval}) and non-magnetic showing intermediate
valent (CeFeSi) behavior \cite{Welter}. Thus, this system allows
to run through a critical region where magnetic order vanishes.

\section{Experimental results}

\subsection{Structural properties}

\begin{figure}
\begin{center}
\includegraphics[angle=0, width=0.45 \textwidth, height=1.2\linewidth] {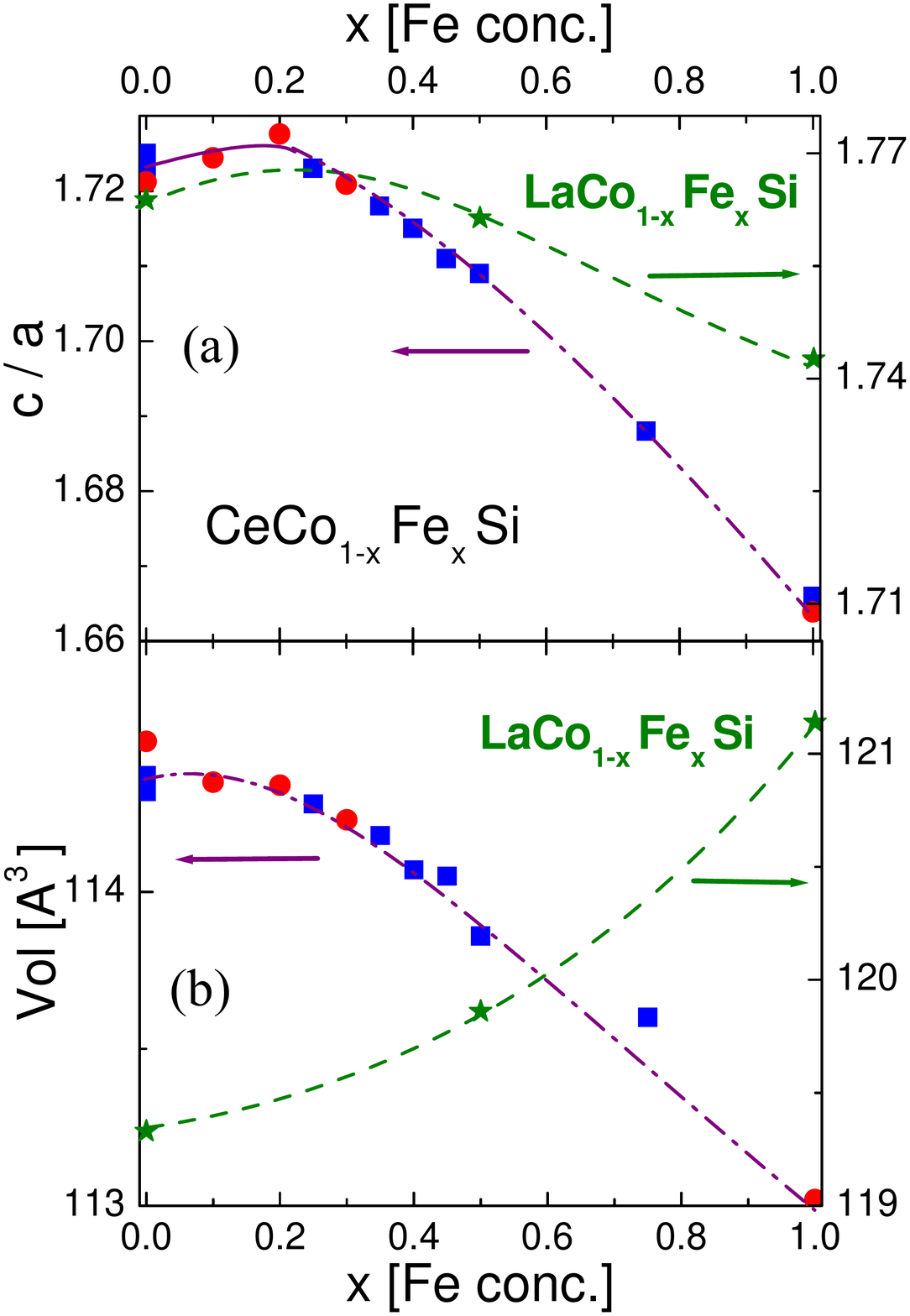}
\end{center}
\caption{(a) c/a lattice parameters ratio and (b) unit cell volume
variation as a function of Fe content in CeCo$_{1-x}$Fe$_x$Si
(left axis). Respective reference parameters form
LaCo$_{1-x}$Fe$_x$Si are included (green) using the right axis.
Dash-dot and dash curves are guide for the eye.} \label{F1}
\end{figure}

The relevant interatomic distances in the CeFeSi-type structure
are the following: $d_{T-X}, d_{R-T}$ and $d_{R-X} $, being R the
rare earth (Ce in this case), T the transition metal (Co/Fe) and X
the semi-metal (Si). The crystal chemistry study performed on the
isotypic compound LaFeSi \cite{Welter2} indicates a reduction of
the mentioned distances respect to the corresponding values of
respective pure elements: $\Delta_{T-Si}= [d_{T-Si} - (d_{T-T}+
d_{Si-Si})]/[(d_{T-T}+ d_{Si-Si})]*100 = -0.084\%$, $\Delta_{R-T}=
-0.027\%$ and $\Delta_{R-Si}= -0.024\%$, whereas the other
distances: $\Delta_{T-T}, \Delta_{R-X}$ and $\Delta_{R-R}$
increase. The large contraction observed on $\Delta_{T-Si}$ is
related to the strong electronic hybridization between those
atoms. This explains the non-magnetic behavior of Fe (or Co) atoms
as a consequence of a large band width.

In the system under study, two regions can be clearly
distinguished in the concentration dependence of the lattice
parameters, see Fig.~\ref{F1}. On the Co-rich side there are minor
variations up to about $20\%$ of Fe doping. Beyond that
concentration, a clear modification occurs in the variation of
both tetragonal axis: a(x) increasing about $1\%$ whilst c(x)
decreases about $5\%$ once it reaches the CeFeSi stoichiometric
limit. Fig.~\ref{F1}(a) shows a significant reduction of the c/a
ratio (about $3.5\%$), coincident with a decrease of the unit cell
volume (about $1.3\%$), shown in Fig.~\ref{F1}(b). This structural
variation tends to reduce Ce-Ce spacing $d_{Ce-Ce}$ between
neighbors placed on the apex and on the square-base pyramides
formed by the each Ce double layer.

The volume variation above $x\approx 0.20$ (see Fig.~\ref{F1}b)
largely exceeds the equivalent variation for neighboring
lanthanides (i.e. La and Pr) as a sign of the Ce-4f orbitals
instability with Fe content increase. As a reference, the
respective values of three La(Co,Fe)Si samples are included in
Fig.~\ref{F1}(a) and (b) on the right axis. The increase of volume
between LaTSi (T = Co and Fe) compounds contrasts with the strong
decrease in CeCo$_{1-x}$Fe$_x$Si alloys, indicating the collapse
of Ce atomic volume for $x\geq 0.20$. In fact, the CeCoSi unit
cell volume is $\approx 1\%$ below the interpolation between
LaCoSi and PrCoSi, whereas CeFeSi is $\approx 4\%$ below the
equivalent La to Pr interpolation.

\begin{figure}
\begin{center}
\includegraphics[angle=0, width=0.5 \textwidth, height=1.4\linewidth] {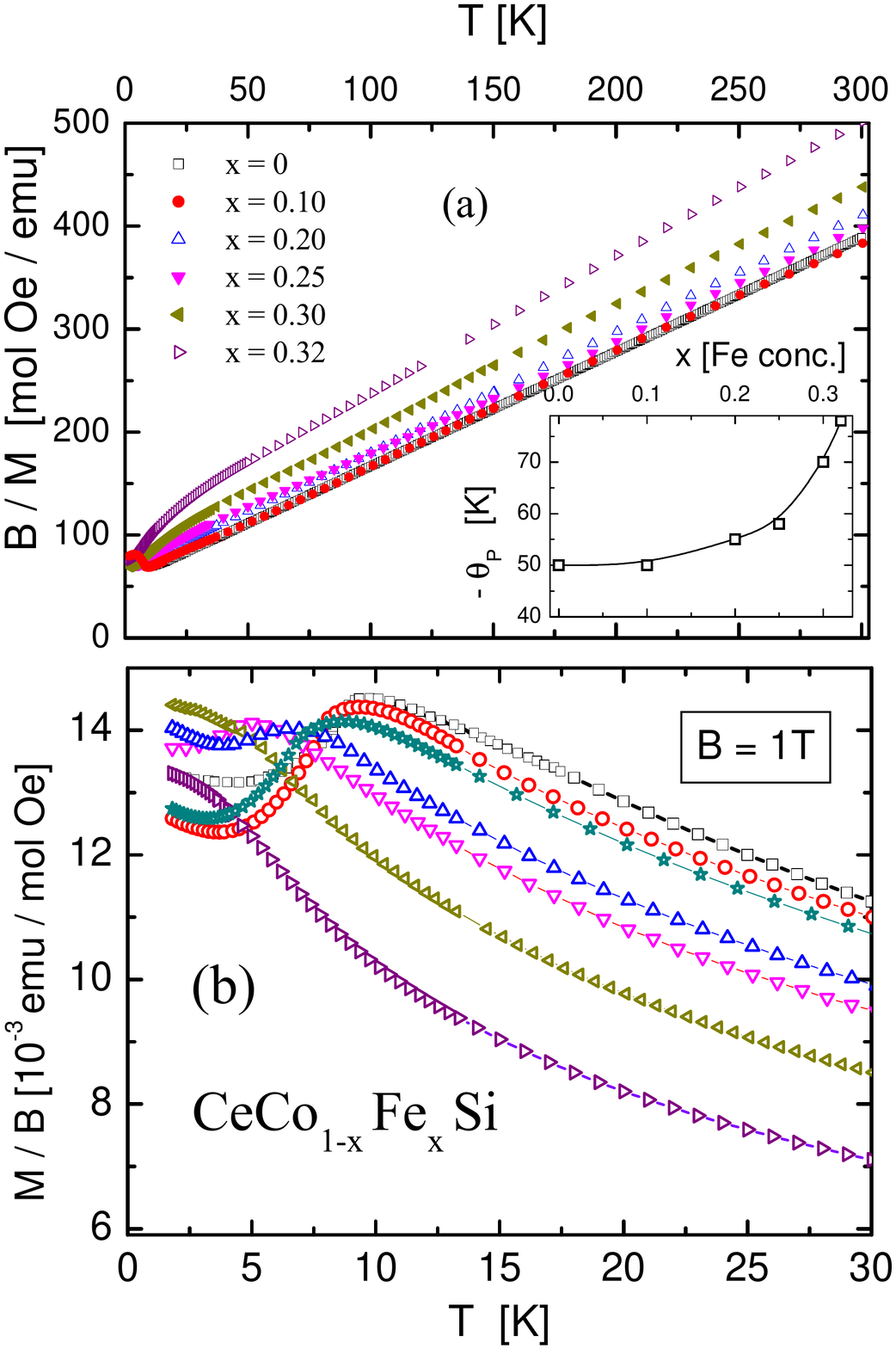}
\end{center}
\caption{(a) High temperature inverse of magnetic susceptibility
for $0\leq x \leq 0.32$. Data of CeCoSi were extracted from Ref.
\cite{Cheval}. Inset: paramagnetic temperature $\theta_P$ as a
function of Fe concentration extrapolated from $T<50$\,K. (b) Low
temperature susceptibility showing the shift of the maximum
connected with the antiferromagnetric order at low temperature
with increasing Fe content.} \label{F2}
\end{figure}

\subsection{Magnetic properties}

The inverse of the high temperature magnetic susceptibility is
shown in Fig.~\ref{F2}(a), from which a Curie constant close to
the $J=5/2$ Hund's rule GS value is extracted from $T>50$\,K for
the $x \leq 0.10$ alloys. Beyond that Fe concentration, the Curie
constant starts to decrease in coincidence with the increase of
the paramagnetic temperature $\theta_P(x)$ as presented in the
inset of Fig.~\ref{F2}(a). Higher Fe concentration samples show
low temperature ferromagnetic contributions that become important
as the main magnetization decreases. Whether this is an intrinsic
effect or due to a foreign contribution is under investigation.

The relatively high value of $\theta_P$ at low Fe content is
likely due to a combination of Kondo effect acting on respective
CEF levels and RKKY exchange. The up turn of $\theta_P(x)$ beyond
$x=0.20$ is due to a strong increase of Kondo screening related to
the significant volume reduction. The corresponding Kondo
temperature ($T_K$) is evaluated following Krishna-Murthy
criterion: $T_K = - \theta_P /2$ \cite{Krishna}.

Details of the low temperature magnetic behavior are presented
Fig.~\ref{F2}(b), showing the decrease of the maximum of $M(T)$ at
the anti-ferromagnetic ordering from $T_N = 9$\,K at $x=0$ down to
$\approx 2$\,K at $x=0.30$. The later is hardly seen in
Fig.~\ref{F2}b because it contains curves performed with an
applied magnetic field of $B=1$\,T, albeit the maximum is clearly
observed in lower applied field e.g. $B = 0.03$\,T (not shown).

\begin{figure}
\begin{center}
\includegraphics[angle=0,width=0.5\textwidth,height=1.4\linewidth] {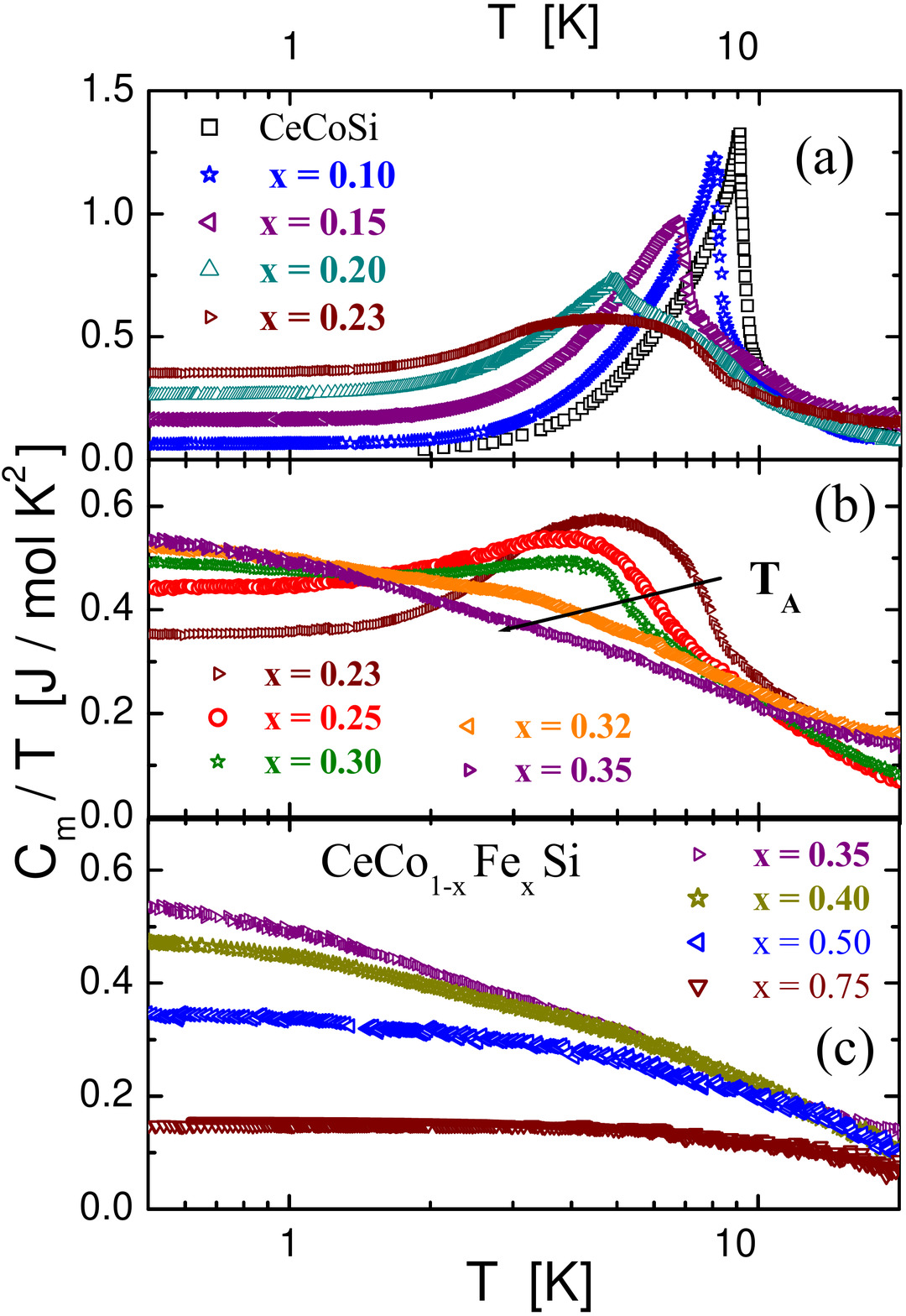}
\end{center}
\caption{Specific heat divided by temperature of all studied
samples in a logarithmic representation. (a) Co-rich side ($0 \leq
x \leq 0.20$) showing the decreasing specific heat jump at the
magnetic transition. (b) $0.23  \leq x \leq 032$ range where the
$C_A$ anomaly becomes dominant (see the text). The arrow indicates
the temperature $T_A$ at the maximum $\partial C_A/\partial T$
slope. (c) The non-magnetic region for $x\geq 0.35$.} \label{F3}
\end{figure}

\subsection{Specific heat}

The magnetic contribution to the specific heat ($C_m$) is obtained
after subtracting the phonon contribution extracted from the
isotypic La compound. The results of all studied samples are
collected in Fig.~\ref{F3}. Three distinct behaviors are observed
in different concentration ranges: (a) on the Co-rich side with a
rapidly decrease of the $C_m(T_N)$ jump $\Delta C_m(T_N)$ within
the $0 \leq x \leq 0.20$ concentration range, (b) a broad anomaly
($C_A$) that becomes dominant between $x=0.23$ and 0.32, and (c) a
non-magnetic region for $x\geq 0.35$. Although these three regions
are clearly distinguishable, the onset of the $C_A(T)$ anomaly is
already detected at lower Fe concentration right above the
$T_N(x)$ transition. Notably, this incipient anomaly starts to
develop at $T\approx 10 \approx T_N(x=0)$\,K. This fact indicates
that the anomaly builds up from the same degrees of freedom but
governed by short range magnetic interactions.

As the $\Delta C_m(T_N)/T$ jump weakens and the $C_A$ anomaly
becomes dominant, also an underlying Kondo type contribution
arises. This contribution is related to the formation of heavy
fermion (HF) quasi-particles. The low temperature value of this HF
contribution increases continuously with $x$ as indicated by the
$C_m/T\mid_{\lim T \to 0}(x)$ values. At $x = 0.35$, a
non-Fermi-liquid (NFL) type dependence: $C_m/T \propto - log(T)$
is clearly seen, although it tends to flatten at low temperature.
For $x > 0.5$ a progressive transformation into a Fermi-liquid
regime occurs as the Ce-lattice enters into a valence instability
regime concomitant with the Ce-volume reduction.

\begin{figure}
\begin{center}
\includegraphics[angle=0, width=0.5 \textwidth] {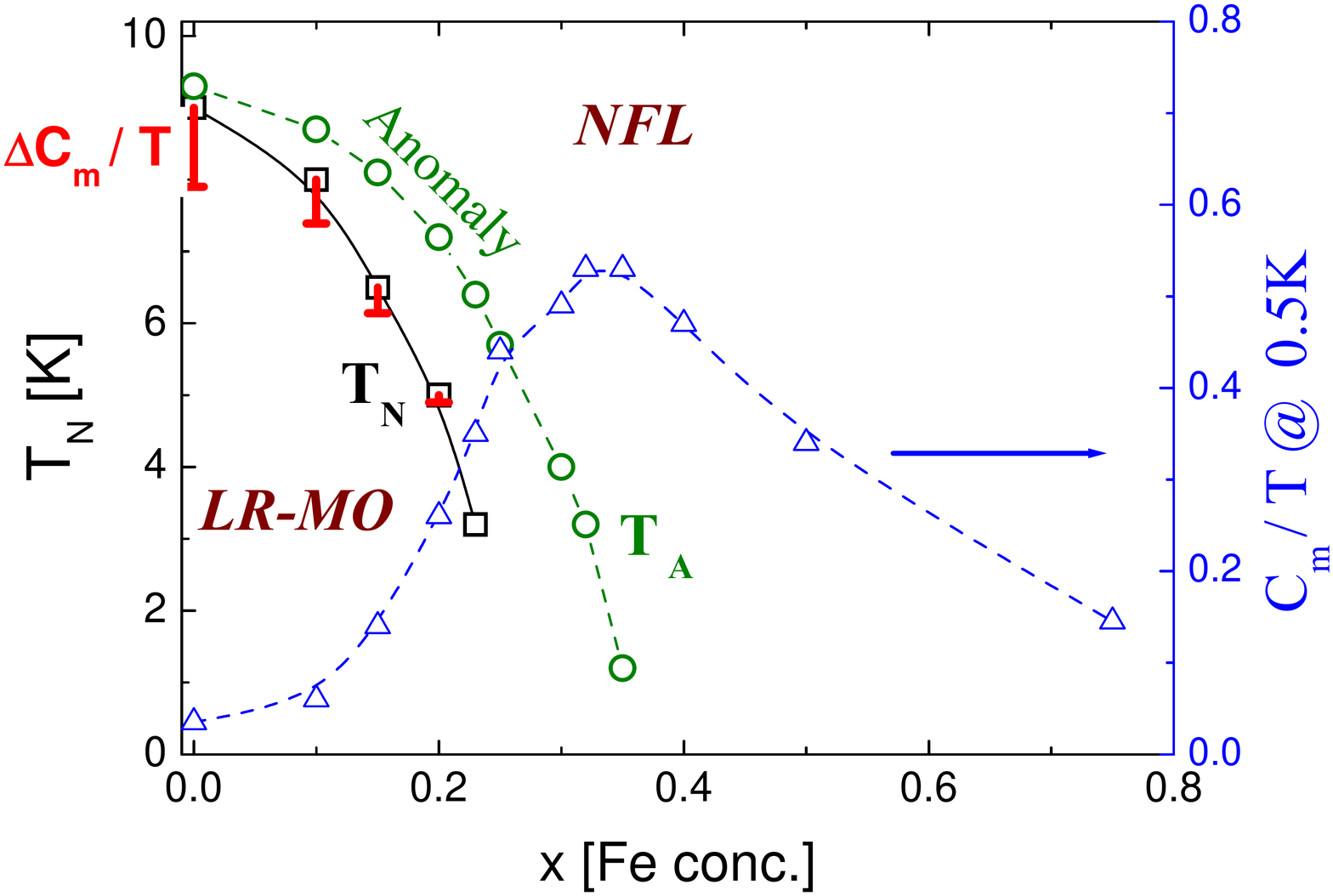}
\end{center}
\caption{Phase diagram collecting ($\Box$) $T_N(x)$ with (red)
bars representing the decreasing jump $\Delta C_m(T_N)$. The lower
$T_N(x=0.23)$ point indicates the temperature of the deviation
from the AF behavior as $\Delta C_m(T_N) \to 0$. ($\circ$)
$T_A(x)$ represents the characteristic temperature of the onset of
the broad anomaly, defined in Fig.~\ref{F3}b. ($\bigtriangleup$)
$C_m/T(x)$ values at 0.5\,K.} \label{F4}
\end{figure}

The temperature of the specific heat jump at $T_N$ does not
coincide with the maximum of $M(T)$ but with the maximum of its
$\partial M/\partial T$ derivative. This is a typical sign of low
dimensionality or strong anisotropy character of magnetic
interactions \cite{Jongh}.

\section{Discussion}

\subsection{Phase diagram}

The concentration dependence of the relevant parameters is resumed
in a schematic phase diagram in Fig.~\ref{F4}. It includes the
$T_N(x)$ decrease together with the respective specific heat jumps
$\Delta C_m(T_N)/T$ reduction that vanishes at $x=0.23$. The
decreasing height of $\Delta C_m/T$ at $T=T_N(x)$ is qualitatively
represented in the figure as a segment attached to each point. The
$C_A(T)$ anomaly is represented by the characteristic temperature
$T_A(x)$ defined in Fig.~\ref{F3}(b) as the temperature of the
maximum negative $\partial C_A/\partial T$ slope above $T_N(x)$.
The HF component is represented by the $C_m/T$ values at
$T=0.5$\,K, which increases in the region $ 0 < x < 0.35$ where
degrees of freedom are progressively transferred from the ordered
phase into the heavy quasiparticles. Beyond that concentration,
$C_m/T$ decreases with $x$, since $C_m/T \propto 1/T_K$ the Kondo
temperature increases as expected for a non-ordered Kondo lattice.

\subsection{On the nature of the specific heat anomaly}

\begin{figure}
\begin{center}
\includegraphics[angle=0, width=0.45 \textwidth] {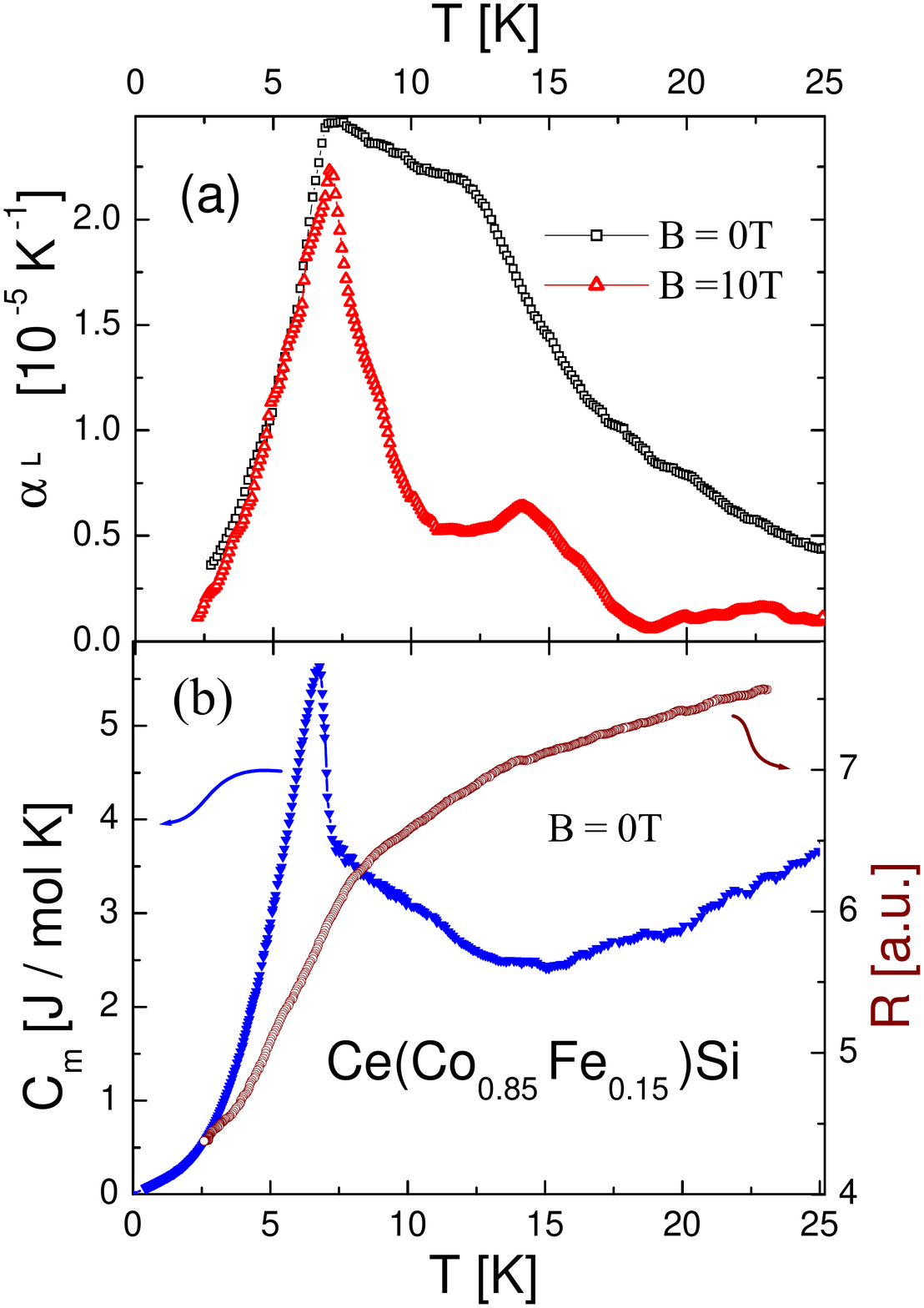}
\end{center}
\caption{(a) Thermal expansion at zero and high magnetic field in
the longitudinal configuration (L//B) of sample $x = 0.15$. (b)
Comparison with zero field specific heat and electrical resistance
(right axis).} \label{F5}
\end{figure}

The origin and nature of the $C_A(T)$ anomaly deserves a more
detailed analysis. As mentioned before, it emerges like a
$T>T_N(x)$ tail as soon as $T_N(x)$ starts to decrease and is
fully developed once the LR-MO phase is suppressed before the
Kondo effects starts to dominate the low temperature behavior. We
notice that the temperature dependence $C_m \propto T^2$, observed
below $T_N(x)$ between $x=0$ and $x=0.35$, remains unchanged even
once the $\Delta C_m/T_N$ vanishes at $x=0.23$. These features
indicate that the $C_A(T)$ anomaly builds up from the magnetic
degrees of freedom which at lower Fe concentration are involved in
the LR-MO phase formation. This anomaly very likely corresponds to
short range magnetic correlations of quasi two-dimensional (2D)
type within the double Ce layer. The increasing Kondo interaction
and the atomic disorder in the Co/Fe-Si layer in between very
likely results in a weakening of the magnetic interactions along
'c' direction and then to a transition from a 3D to a quasi 2D-
system.

In order to gain insight into the characteristics of this anomaly,
we have performed thermal expansion ($\alpha$) measurements at
zero and under strong magnetic field ($B = 10$\,T) on sample
$x=0.15$. At that Fe concentration both, the incipient $C_A$
anomaly and the $T_N$ transition are competing, see
Fig.~\ref{F5}(a). For comparison, zero field specific heat and
electrical resistance ($R$) are included in Fig.~\ref{F5}(b). The
temperature dependence of these three properties coincide in
showing the onset of the anomaly contribution at $T\approx
12.5$\,K, being $\alpha(T)$ the more sensitive to the appearance
of those magnetic correlations. Incipient coherence effects are
observed in the $R(T)$ dependence starting at $T \approx 13$\,K.
This temperature coincides with the onset of the $C_m(T)$ upturn
approaching the magnetic transition.

Notably, applied magnetic field of $B=4$\,T (not shown) produces
no relevant effects neither on the anomaly nor on the $T_N$
transition. A $B = 10$\,T field weakens the anomaly without
affecting the $T_N$ transition (see Fig.~\ref{F5}a) like in
stoichiometric CeCoSi \cite{Cheval2}. Furthermore, a slight shift
of the transition to lower temperature (to $T_N\approx 5$\,K) is
only observed by applying a field $B=16$\,T. Preliminary $M(B)$
measurements up to $B=5$\,T show a very weak upturn from a linear
dependence for $B\geq 4$\,T, with coincident values obtained at
$T=1.8, 3$ and 5\,K.

\section{Summary}

Although the magnetic order of this system vanishes as expected,
the phase boundary of the long range magnetic order does not
extrapolate to $T=0$. Instead we observe a progressive
substitution of the LR-MO by short range interactions. A broad
anomaly arises between the $T_N(x=0)$ temperature and the actual
decreasing $\Delta C_m(T_N(x))$ jump of each alloy. These features
suggest that the involved magnetic degrees of freedom are
transferred from one component to the other. Notably, the magnetic
interactions in this system are very robust against external
magnetic field application because both, $C_A$ anomaly and the
$\Delta C(T_N)$ jump, require $B\geq 10$\,T and $B\geq 14$\,T to
be suppressed or shifted respectively.

Beyond the $C_A$ phase formation, the complex phase diagram shows
how the magnetic degrees of freedom from the Co-rich side are
transferred to a heavy quasiparticles component exhibiting a NFL
type behavior. Once the full NFL regime is reached around $x
\approx 0.35$, $T_K$ increases rapidly towards the Fe-rich side.
The large value of $\theta_P$ and the entropy gain with
temperature indicates that the first excited CEF level is also
affected by the Kondo effect and partially contributes to the low
temperature properties. Further studies are in progress to better
elucidate the exotic characteristics of the $C_A$ anomaly and the
significant magnetic hardness of these alloys.

This investigation confirms that Ce-equiatomic ternary compounds
with strongly anisotropic structures allows to access to novel
behaviors where enhanced fluctuations can play an important role.

\end{document}